\documentclass[11pt,a4paper]{article}
\usepackage[utf8]{inputenc}
\usepackage{amsmath}
\usepackage{amsfonts}
\usepackage{mathtools}
\usepackage{amssymb}
\usepackage{color}
\usepackage{cancel}
\usepackage[top=2.0cm, bottom=2.0cm, left=2.5cm, right=2.5cm]{geometry}
\usepackage{graphicx}
\usepackage{natbib}
\usepackage{setspace} 
\usepackage{calligra}
\usepackage{calrsfs}
\usepackage{lineno}

\newcommand{\ua}{{\bf u}_{{A}}}
\newcommand{\uh}{{\bf u}}

\DeclareMathAlphabet{\mathcalligra}{T1}{calligra}{m}{n}
\DeclareMathAlphabet{\pazocal}{OMS}{zplm}{m}{n}

\title{Alfv\'en Waves at Mars}
\date{}
\author{\hspace{-1.5cm}
N. Romanelli$^{1,2,*}$, C. M. Fowler$^{3}$,  G. A. DiBraccio$^{2}$,  J. R. Espley$^{2}$, J. S. Halekas$^{4}$\vspace{.3cm}\\ \hspace{-1.5cm}
(1) Department of
Astronomy, University of Maryland, College Park, MD, USA.\\\hspace{-2cm} (2) Planetary Magnetospheres Laboratory, NASA Goddard Space Flight Center, Greenbelt, MD, USA.\\\hspace{-2.0cm}
(3) Department of Physics and Astronomy, West Virginia University, Morgantown, WV, USA.\\\hspace{-1.5cm}
(4) Department of Physics and Astronomy, University of Iowa, Iowa City, IA, USA. \vspace{.3cm}\\\hspace{-1.5cm}
(*) Corresponding author: norberto.romanelli@nasa.gov
}

\begin{document}

\maketitle 

\doublespacing

\section*{Abstract}
The solar wind upstream of Mars's bow shock can be described in terms of Alfv\'enic turbulence, with an incompressible energy cascade rate of $10^{-17}$ J m$^{-3}$ s$^{-1}$ at magnetohydrodynamics (MHD) scales. The solar wind has more Alfv\'en waves propagating outwards from the Sun (than inwards) and a median Alfv\'en ratio of $\sim0.33$.
Newly ionized planetary protons associated with the extended hydrogen corona generate waves at the local proton cyclotron frequency. These 'proton cyclotron waves' (PCW) mostly correspond to fast magnetosonic waves, although the ion cyclotron (Alfv\'enic) wave mode is possible for large Interplanetary Magnetic Field cone angles. PCW do not show significant effects on the solar wind energy cascade rates at MHD scales but could affect smaller scales. The magnetosheath displays high amplitude wave activity, with high occurrence rate of Alfv\'en waves. Turbulence appears not fully developed in the magnetosheath, suggesting fluctuations do not have enough time to interact in this small-size region. Some studies suggest PCW affect turbulence in the magnetosheath. Overall, wave activity is reduced inside the magnetic pile-up region and the Martian ionosphere. However, under certain conditions, upstream waves can reach the upper ionosphere. So far, there have not been conclusive observations of Alfv\'en waves in the ionosphere or along crustal magnetic fields, which could be due to the lack of adequate observations.

\section{Introduction}

More than eighty years ago, {Hannes} Alfv\'en proposed the existence of a new type of electromagnetic-hydrodynamic wave capable of propagating in a conductive fluid medium \citep{Alfven1942}. Currently known as Alfv\'en waves, they are the result of the coupling between the plasma motion and magnetic field line stresses. Plasma flow across magnetic field lines bends them, perturbing the current density field, modifying the total magnetic field force, and ultimately affecting plasma motion itself. {Alfv\'en waves, 
 fast and slow magnetosonic waves, are the three low-frequency normal modes present in a magnetized ideal conductive fluid medium, whose evolution over long time scales, is described by means of magnetohydrodynamics or MHD theory} \citep[e.g.,][]{Cramer2001}. {In the MHD regime,} the Alfv\'en wave mode is incompressible, not presenting perturbations in the plasma density and particle pressure. The magnetic field strength is also unaffected by this normal wave mode. Identifying wave modes is {important} as, under many conditions, they constitute a non-collisional coupling in space and astrophysical plasmas. Moreover, this identification also allows us to indirectly infer other physical processes occurring in a variety of environments, such as the solar wind and the {magnetospheres} of planets \citep[e.g.,][]{KivelsonRussell1995}.  

The pristine solar wind upstream from Mars has been studied in terms of incompressible Alfv\'enic turbulence {\citep{Parker1958,Po2001}}. Nonlinear wave activity {is modeled in terms of}  the interaction of counter-streaming Alfv\'en waves and by the presence of an inertial range, where energy is transferred without dissipation throughout several spatial and temporal scales \citep[e.g.,][]{FR1995,Bruno2013,Andres2020}. In the MHD scales, the magnetic field power spectral density displays a Kolmogorov-like form. {In this regime, the power of magnetic field fluctuations follows a power law with the wave frequency, and the spectral index is equal to -5/3 \citep[e.g.,][]{Alexandrova2008,Ruhunusiri2017,Andres2020}. } 

{The interaction between the magnetized solar wind and the Martian ionosphere contributes to a planetary magnetosphere that bears similarities with those of Venus and comets \citep[e.g.,][]{Acuna1998,Acuna1999,Mazelle2004,Nagy2004,Halekas2021,Cravens2004,Bertucci2011,Dubinin2023}. On the other hand, the presence of crustal magnetic fields and the interaction with the Interplanetary Magnetic Field (IMF) is responsible for features also present in intrinsically magnetized planetary magnetospheres.}
 In particular, crustal magnetic fields create mini-magnetospheres, with characteristics that resemble those of magnetized planets but at localized scales; they can magnetically reconnect with the solar wind IMF, providing the solar wind access to localized regions of the ionosphere (in analogy to the cusp regions at Earth); {and contribute to the dynamics of the environment.} {Indeed, crustal magnetic fields rotate with the planet}, adding to the proper solar wind variability 
\citep[e.g.,][]{Acuna1998,Acuna1999,Brain2003,Harada2018,DiBraccio2022,Bowers2023}. {As a result,} the interaction between the solar wind/IMF, the Martian ionosphere, and the remanent crustal magnetic fields results in a hybrid planetary magnetosphere \citep[e.g.,][]{DiBraccio2022,Dubinin2023}.

\begin{figure}[ht]
\centering
\includegraphics[scale=1.2]{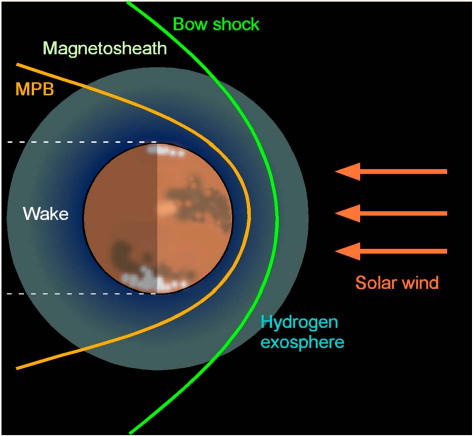}
\caption{Schematic representation of the Martian magnetosphere. The bow shock is the boundary that separates the upstream solar wind from the magnetosheath region. It contributes to solar wind heating and deceleration. The magnetic pileup boundary (or MPB) is the inner boundary that separates the magnetosheath from the magnetic pileup region. Downstream of the MPB, the plasma is mostly composed of planetary heavy ions. Due to Mars' relatively weak gravity field and the lack of an intrinsic global magnetic field, the hydrogen corona extends beyond Mars's bow shock. Source: Taken from \cite{Ruhunusiri2017}.}
 \label{Fig_1_Ruh_2017}
\end{figure}

The first boundary the solar wind encounters when interacting with the Martian magnetosphere is the bow shock, {as shown in Figure \ref{Fig_1_Ruh_2017}}. Electromagnetic forces in this collisionless boundary slow down and heat the incoming supermagnetosonic solar wind.
As the upstream solar wind Alfv\'en Mach number is high at Mars heliocentric distances ($M_A \sim 11$), backstreaming protons are generated, which contribute to diverting the solar wind around the planet \citep{Halekas2017,Slavin1981,Edmiston1984,Paschmann1980,Sonnerup1969,Biskamp1973,Gosling1985,Phillips1972}. Thus, a foreshock is present at Mars under nominal solar wind conditions \citep{Eastwood2005,Jarvinen2022,Burgess2005,Bale2005}.  The backstreaming protons interact with the incoming solar wind and give rise to low-frequency waves immersed in the Martian foreshock \citep[e.g.,][]{Meziane2017,R2018c,Jarvinen2022}. In addition, the shock  generates electromagnetic plasma waves capable of overcoming the solar wind velocity and traveling upstream \citep[e.g.,][]{Mazelle2004}.
The average bow shock stand-off distance is $\sim 1.6 R_M$, where $1 R_M\sim 3390$ km, significantly smaller relative to those associated with intrinsically magnetized planets \citep[e.g.,][]{Mazelle2004,Gruesbeck2018,Eastwood2005,Turc2023}. {For instance, the average Earth's bow shock stand-off distance is  $\sim 14 R_E$, where $R_E$ stands for Earth's radius ($1 R_E\sim 6378$ km) \citep[e.g.,][]{Fairfield1971,Formisano1979}.}

The region downstream of the bow shock, the magnetosheath, is characterized by compressed, heated, and slower solar wind plasma and high-amplitude wave activity \citep[e.g.,][]{Halekas2017,Halekas2017seas,Fowler2017}. The sources of these waves are numerous, from transmission through the bow shock, to local generation to energy transfer between modes and scales \citep[e.g.,][]{Dubinin2016,Ruhunusiri2017}. The inner boundary of the magnetosheath is the magnetic pile-up boundary or MPB, whose average stand-off distance is $\sim 1.2\, R_M$. {Dowstream from the MPB,  the background magnetic field drapes and piles up around Mars \citep[e.g.,][]{Trotignon2006,Bertucci2011,Romanelli2014,Espley2018}. Moreover,} the plasma is mostly composed of planetary heavy ions and the magnetic field strength is generally larger than in the magnetosheath, resulting in a region of relatively dense and low plasma beta and generally less intense wave activity \citep{Nagy2004,Fowler2017}. 
Additionally, Mars also has a magnetotail consisting of magnetic lobes of opposite polarity separated by a current sheet whose location depends on the IMF \citep[e.g.,][]{Crider2004,Romanelli2015,Romanelli2018_CF,Harada2015,DiBraccio2017,DiBraccio2018,DiBraccio2022,Ramstad2020,Curry2022}. {
A significant fraction of the total planetary ion escape occurs down the magnetotail region, with particle fluxes dependent on position, particle energy, and external conditions (such as the orientation of the IMF), among other parameters \citep[e.g.,][]{Dubinin2011,Curry2022}.}

 Mars also has a hydrogen (H) exosphere that extends beyond the bow shock \citep{Bhattacharyya2015,Chaffin2015}. This is due to Mars's relatively weak gravity and small size magnetosphere due, in turn, to the lack of a global intrinsic magnetic field \citep{Acuna1998}. The neutral exospheric atoms are ionized by photoionization, charge exchange, and electron impact, and give rise to a newborn planetary proton  population, initially nearly at rest with respect to Mars \citep{Yamauchi2015,Rahmati2015,Rahmati2017}. Seasonal variability of the H corona results in seasonal variability of the newborn planetary proton population, also creating seasonal variability in physical processes occurring inside and upstream of the magnetosphere of Mars. {In particular, previous studies have reported seasonal and long-term variability} of low-frequency wave activity upstream of Mars and effects inside the magnetosphere \citep[e.g.,][]{Bhattacharyya2015,Yamauchi2015,Ruhunusiri2017,Halekas2017seas,Romanelli2016,Romeo2021,Hughes2019,Halekas2020,Jiang2023}.

{In addition, the relatively small scale size of the Martian magnetosphere means that characteristic length scales of the plasma (such as the proton gyro radius) are comparable to the length scales of the system (e.g. bow shock stand-off distance), implying that the shocked solar wind does not have space to fully thermalize before encountering the planet. Kinetic effects are thus likely to play a key role in the Mars - solar wind interaction  \citep[e.g.,][]{Moses1988,Kallio2011,Dong2015,Romanelli2019,Jarvinen2022}.}
  In particular, although some of the observed wave properties may be understood in terms of MHD normal modes, previous analyses have shown statistical and case studies where wave polarization and frequency and the local ion and electron velocity distribution functions necessarily demand a kinetic description. This is the case, for instance, for most ultra-low frequency (ULF) waves observed upstream from the Martian bow shock and in the foreshock. The observed polarization is typically elliptical or circular, while MHD predicts linearly polarized modes. Moreover, particle populations in these regions are far from Maxwellians \citep[e.g.,][]{Russell1994,Mazelle2004,Romanelli2016,Meziane2017,Romeo2021,Jarvinen2022}.
It is also worth mentioning that the presence of boundaries (bow shock and MPB) affects the possible wave modes excited in the Martian magnetosphere. {Examples of associated processes are mode conversion, wave reflection as well as surface waves, among others.}

The relatively small size of the hybrid Martian magnetosphere also suggests that waves resulting from {the interaction} with the solar wind should be present downstream of the bow shock and even in the ionosphere. Subsequent observations have confirmed that this is the case and the magnetosphere of Mars is now known to be a highly dynamic environment, where processes taking place upstream from the bow shock can significantly impact the underlying ionosphere \citep[e.g.,][]{Halekas2017,Fowler2017,Ruhunusiri2017,Jiang2023}. 
Studying the electromagnetic wave environment at Mars is thus very important as {waves} can facilitate energy propagation and transfer in the collisionless environment, heat the ionosphere and potentially modify planetary ion escape \citep[e.g.,][]{Ergun2006,Fowler2018,Collinson2018,Jarvinen2022}.  Furthermore, {waves} can lead or modify magnetic field reconnection conditions in the Martian magnetosphere \citep{Chen2022,Bowers2023}.
In other words, the presence of an extended H exosphere, the relatively small size of the Martian magnetosphere, and the remanent crustal magnetic fields, among other parameters, {provide a compelling} plasma environment in the solar system to study electromagnetic plasma wave generation and evolution. 

Different planetary missions to Mars {have} allowed the identification of several low-frequency waves. The first observations of wave activity at Mars were obtained by the Phobos‐2 mission \citep[e.g.,][]{Grard1989,Russell1990,Skalsky1998,Delva1998}.
Since then, analysis of magnetic, electric, and plasma observations by Phobos-2, Mars Global Surveyor (MGS), Mars EXpress (MEX), and Mars Atmosphere and Volatile EvolutioN (MAVEN) missions to Mars presented us with a picture where certain types of waves are more likely to be observed in different regions of the Martian magnetosphere and upstream, with variability in their properties and occurrence rate. Despite this progress, comprehensive analysis of low-frequency waves in the Martian magnetosphere has been limited for several reasons. The Phobos-2 mission {had a
relatively high periapsis ($\sim$ 900 km) and lasted} $\sim 8.5$ months, resulting in a relatively small amount of data. MGS lacked an ion instrument capable of measuring the bulk properties of local plasma. Additionally, most of its orbits were Sun-synchronous at $\sim 400$ km, affecting the spatial coverage of the Martian magnetosphere throughout this mission (11 September 1997 - 2 November 2006). Since 25 December 2003, MEX has explored the Martian magnetosphere,  providing valuable measurements that continue to improve our understanding of plasma processes. However, the lack of a magnetometer onboard MEX puts constraints on the characterization of wave modes around Mars.  Inserted into orbit around Mars on 21 September 2014, MAVEN is the first and only spacecraft so far to observe {all main regions of} the magnetosphere carrying a magnetometer, an ion plasma analyzer, and an electric field instrument. Moreover, its precessing orbit  {also allows sampling of the entire magnetosphere} under different solar wind and solar cycle conditions \citep{Jakosky2015}.

In this Chapter, we provide a review focused on Alfv\'en waves, taking into account the presence of other wave modes. Several excellent reviews on low-frequency waves at Mars provide a comprehensive context for this work \citep{Russell1994,Mazelle2004,Glassmeier2006,Delva2011,Dubinin2016}.
This Chapter is centered on and highlights some of the many recent observations and conclusions provided by the ongoing MAVEN mission \citep{Jakosky2015}.

\section{Alfv\'en waves upstream from the Martian Bow Shock}

Alfv\'en waves of several wavelengths are present in the upstream solar wind. These waves are responsible for correlated fluctuations in the magnetic and velocity fields, which can affect the Martian plasma environment \citep{Belcher1971,Halekas2017,Andres2020}. In addition, the nonlinear interaction between counter-streaming Alfv\'en waves is considered to be responsible for the energy cascade, by which energy is transferred throughout several scales of the system. Under many conditions, the sense of the nonlinear energy transfer is directed from the larger wavelengths (or timescales) to the smaller ones. This is the so-called direct incompressible energy cascade of the solar wind \citep{Bruno2013,Alexandrova2008,Maron2001,Ruhunusiri2017,Andres2020,Romanelli2022}. 

A study conducted by \cite{Halekas2017} made use of 45 s average MAVEN Magnetometer (MAG) and Solar Wind Ion Analyzer (SWIA) observations to study the Alfv\'enic content in the Martian magnetosphere and upstream of the bow shock \citep{Connerney2015ssr,Halekas2015}. Among the  explored properties, the authors computed the normalized cross-helicity $\sigma_C$, the normalized residual energy $\sigma_r$, and the Alfv\'en ratio, defined as follows: 

\begin{equation}
    \sigma_C =  \frac{2\,\langle\delta \mathbf u \cdot \delta\ua\rangle}{\langle \delta \mathbf u^2 + \delta\ua^2\rangle} 
\end{equation}
\begin{equation}
    \sigma_r = \frac{\langle \delta \mathbf u^2 - \delta\ua^2\rangle}{\langle \delta \mathbf u^2 + \delta\ua^2\rangle} 
\end{equation}

\begin{equation}
    r_A =  \frac{\langle\delta \mathbf u^2\rangle}{\langle\delta\ua^2\rangle} 
\end{equation}

where $\delta \mathbf u$ and $\delta\ua$ are the solar wind bulk and Alfv\'en incompressible velocity   fluctuations, respectively. The latter is defined as   $\ua\equiv\textbf{B}/\sqrt{\mu_0 \rho_0}$, where  $\rho_0$ is  the mean mass plasma density and $\mu_0$ is the vacuum magnetic permeability.  The angular bracket $\langle \rangle$ indicates a time average over each 30-minute analyzed interval. Note that, by definition,  
$-1 \leq \sigma_C \leq 1$ , $-1 \leq \sigma_r \leq 1$, and $r_A \geq 0.$ 

The  normalized cross-helicity $\sigma_C$ is a measure of the linear correlation between velocity and magnetic field fluctuations.  A value of $\sigma_C=\pm1$ is consistent with an Alfv\'en wave
propagating antiparallel or parallel to the background magnetic field direction. Usually, fluctuations with 
$|\sigma_C|\sim 1$ are described as Alfv\'enic.
The normalized residual energy and the  
Alfv\'en ratio quantify the energy balance between the kinetic and magnetic field fluctuations. In particular, a value of $\sigma_r=0$ is consistent with energy equipartition, and values of $\sigma_r<0$ show a given event has more energy in the magnetic field fluctuations. {Consistently, a value of $r_A = 0$ implies all energy is present in the magnetic field fluctuations and $r_A = 1$ is associated with energy equipartition.}

\begin{figure}[ht]
\centering
\includegraphics[scale=1.2]{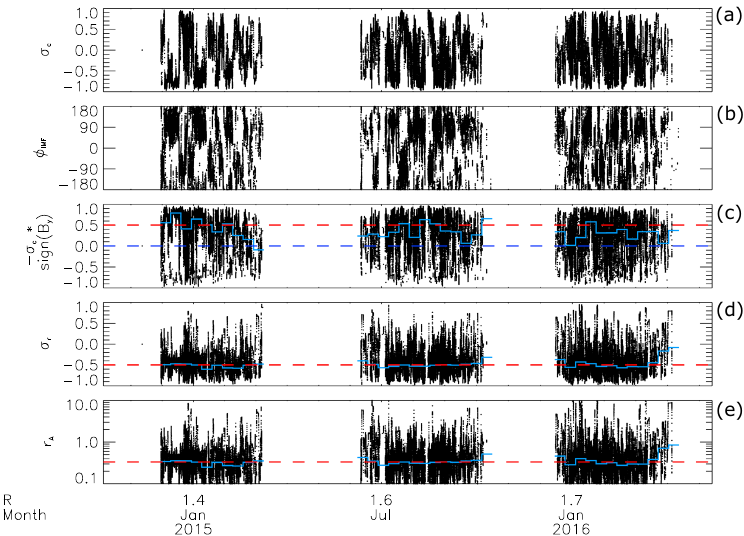}
\caption{Solar wind properties computed from 30-minute time intervals upstream of Mars' bow shock as a function of time, between October 2014
and May 2016 (black points). (a) The normalized cross helicity ($\sigma_C$), (b) azimuthal angle of the IMF in the X-Y MSO plane $\phi_{IMF}=tan^{-1}(By/Bx)$, (c) $-\sigma_C$ multiplied by the sign of the IMF dawn-dusk (By) MSO component, (d) normalized residual energy ($\sigma_r$), (e) and Alfv\'en ratio ($r_A$). The light blue lines in panels (c-e) show the corresponding median values over 10-day intervals. The dashed red lines display the corresponding median values from March to May 2014 (MAVEN's cruise phase) for comparison. Note the dashed blue line in panel (c) is equal to zero and is shown for reference. Text labels indicate Mars's heliocentric distance (R) in astronomical units. MSO: Mars Solar Orbital. IMF: Interplanetary Magnetic Field.  Source: Taken from \cite{Halekas2017}.}
 \label{fig_hal_2017_12}
\end{figure}

\cite{Halekas2017} showed that the solar wind upstream from the Martian bow shock has a majority of Alfv\'en waves that propagate outward from the Sun. Figure \ref{fig_hal_2017_12}a-c displays a larger amount of waves with negative (positive) normalized cross-helicities for +By/-Bx (-By/+Bx) Mars Solar Orbital (MSO) IMF configuration \citep{Roberts1987}. The MSO coordinate system is centered on Mars with the X-axis pointing toward the Sun, and the Z-axis perpendicular to Mars’s orbital plane and positive towards the ecliptic north. The Y-axis completes the
right-handed system.  Moreover, Figure \ref{fig_hal_2017_12}d also shows that solar wind magnetic field fluctuations have more energy than velocity fluctuations, with normalized residual energies on the order of $\sim-0.5$. These results are consistent with previous observations of the solar wind. They provide additional information for the characterization and modeling of the occurrence of Alfv\'en waves  with heliocentric distance, latitude, solar cycle, and solar wind properties \citep{Bruno2013,Bavassano1998,Breech2005,Tu1991}. 

\begin{figure}[ht]
\centering 
\includegraphics[scale=2]{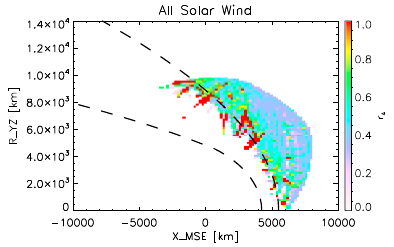}
\caption{The normalized
Alfv\'en ratio for solar wind measurements between October 2014 and May 2016, in cylindrical MSE coordinates. The
dashed black curves display the bow shock and magnetic pile-up fits from \cite{Trotignon2006}. MSE:  Mars Solar Electric.  Source: Taken from \cite{Halekas2017}.}
 \label{fig_hal_2017_13}
\end{figure}

Interestingly, $\sigma_C$ was found to decrease in the quasi-parallel foreshock, possibly associated with effects from nonlinear compressive waves present in this region \citep[see the top and middle row panels, Figure 13 in][]{Halekas2017}. On the other hand, the Alfv\'en ratio was found to increase as the solar wind approaches the Martian bow shock, as shown in Figure \ref{fig_hal_2017_13}, making use of cylindrical Mars Solar Electric (MSE) coordinates. The MSE coordinate system is centered at Mars and its axes are defined as follows. The X-axis point towards the Sun{, including correction 
for solar wind aberration due to Mars' orbital motion}, and the Z-axis point along the convective electric field seen in the planet's rest frame ($\mathbf{E}_{SW}=-\mathbf{U}_{SW} \times \mathbf{B}_{SW} $), where $\mathbf{U}_{SW}$ and 
 $ \mathbf{B}_{SW}$ are the solar wind velocity and IMF, respectively. The Y-axis completes the right-handed system.
These observations constitute another example of  Mars affecting the solar wind,  even upstream from {the bow shock.} The computed Alfv\'en ratio increase, from $\sim 0.2$ to $\sim 1$, is similar for both the quasi-parallel and quasi-perpendicular bow shock regions (not shown), suggesting it is likely not due to foreshock effects. As discussed in \cite{Halekas2017}, the observed energy transfer from magnetic to kinetic fluctuations could be associated with solar wind mass loading by planetary pick-up ions, especially protons resulting from Mars's extended H exosphere \citep{Halekas2017seas,Yamauchi2015,Rahmati2017}.  

Mass-loading is only one of several macroscopic effects associated with cumulative planetary ion pick-up.
Analogous to comet pick-up ion physics, the newly ionized planetary proton population represents an additional, non-thermal component, of the total  proton velocity distribution function \citep[e.g.,][]{Mazelle2004,Gary1993,Delva2011,Delva2015}. This distribution is capable of generating electromagnetic plasma waves upstream from the Martian bow shock, in association with different plasma instabilities \citep{Gary1993}. These waves affect the pristine solar wind before it encounters the bow shock and can be transmitted into the magnetosheath, having also effects over processes in the Martian magnetosphere \citep{Romeo2021,Andres2020,Romanelli2022,Ruhunusiri2017,Ruhunusiri2015,Jiang2023}.

{Several parameters determine which plasma instability has the largest linear wave growth rate.} Among them is the IMF cone angle,  i.e., the angle between the IMF and the solar wind velocity at the time of pick-up.  As shown in \cite{Brincatsu1989}, the ion-ion right-hand resonant (RH) instability is predominant 
when the IMF cone angle is smaller or equal to 75$^\circ{}$. The ion-ion left-hand resonant instability (LH) is the most unstable for IMF cone angles larger than $\sim 75^\circ{}$ \citep{Brincatsu1989}. Both ion-ion resonant instabilities have maximum linear wave growth rates for propagation parallel to the background magnetic field \citep{Brinca1991,Gary1993}. 
Therefore, the presence of pick-up ions is also responsible for the generation of ultra-low frequency waves, which, in the Hall-MHD formalism correspond to the fast magnetosonic (right-hand polarization) and the ion cyclotron waves (left-hand polarization), respectively. Such wave modes tend, in turn, to the MHD fast magnetosonic and the Alfv\'en linearly polarized wave modes for very low frequencies, under parallel propagation conditions \citep{Cramer2001}. Interestingly, the Doppler shift is responsible for the observation of these waves (RH and LH) at a frequency near the local proton cyclotron frequency in the spacecraft frame, with a left-handed elliptical polarization \citep{Gary1989,Brinca1991,Gary1993}. It is because of this particular property that these waves are sometimes referred to as proton cyclotron waves (or PCW). However, as mentioned before, only under very specific conditions {PCW do} they actually correspond to the ion cyclotron (Alfv\'enic) wave branch.  {Note that the nomenclature PCW, is based on wave properties seen in the spacecraft reference frame, while the ion cyclotron wave mode is defined in the plasma rest frame.}

 The first report of PCW upstream from Mars' bow shock was made by \cite{Russell1990}. These ultra-low frequency waves have also been detected and analyzed in greater detail thanks to magnetic field observations provided by MGS and MAVEN missions \citep{Brain2002,Mazelle2004,Bertucci2013,Romanelli2013,Romanelli2016,Liu2020,Wr2006,Wei2011,Wei2014,Romeo2021}. These waves were consistently found left-handed elliptically polarized in the spacecraft reference frame and propagating quasi-parallel to the mean IMF (with propagation angles on the order of 20$^\circ{}$), {which} has been associated with nonlinear effects \citep[e.g.,][]{Romanelli2013}. 
 
\begin{figure}[ht]
\centering
\includegraphics[scale=1.1]{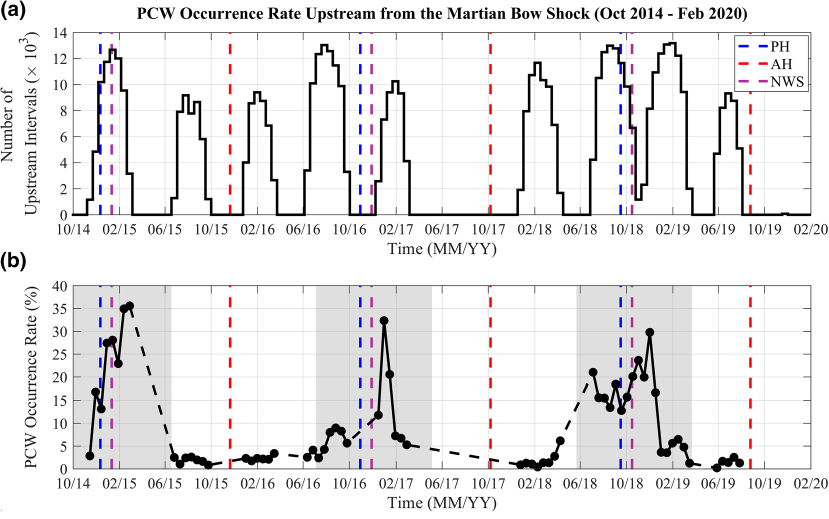}
\caption{(a) Histogram of the number of analyzed 512 s time intervals ($\times 10^3$) upstream from Mars's bow shock as a function of time (15-day bins), between October 2014 and February 2020. (b) Derived PCW occurrence
rate (\%) as a function of time. The dashed black line is associated with data gaps in the upstream region (see upper panel). Gray-shaded areas correspond to periods of dust storm seasonal activity. (a-b) The blue, purple, and red vertical dashed lines correspond to the Martian perihelion (PH), the northern winter solstice (NWS), and the aphelion (AH), respectively. PCW: Proton Cyclotron Waves. Source: \cite{Romeo2021}/ John Wiley \& Sons.}
 \label{fig_romeo_2020}
\end{figure}

 MGS and MAVEN allowed the identification of a long-term trend, with higher PCW occurrence rate near the Martian perihelion  \citep{Romanelli2013,Bertucci2013,Romanelli2016,Romeo2021}. Such an annual trend is likely associated with seasonal changes in the exospheric hydrogen density and resulting newborn planetary proton density \citep{Rahmati2017,Rahmati2018,Halekas2017seas,Yamauchi2015}. Figure \ref{fig_romeo_2020} shows the PCW occurrence rate as a function of time, as seen by MAVEN between October 2014 and February 2020. \cite{Romeo2021} reported that this rate displays an increase up to $\sim 30-35$\% near (slightly after) the Martian perihelion and southern summer solstice. These values are an order of magnitude larger than the average value near the Martian aphelion ($\sim 2$\%), in agreement with previous studies \citep{Romanelli2013,Bertucci2013,Romanelli2016}.
It is also worth noticing that the PCW occurrence rate increase takes place during part of the Martian dust storm season, marked by gray regions in Figure \ref{fig_romeo_2020}b \citep{Romeo2021}. 
{In this context, \cite{Chaffin2021}  reported that a regional dust storm observed well after the perihelion increased planetary H escape by a factor of five to ten, suggesting that dust dynamics affect exospheric H densities more than seasonal variations.}

\begin{figure}[ht]
\centering
\includegraphics[scale=1]{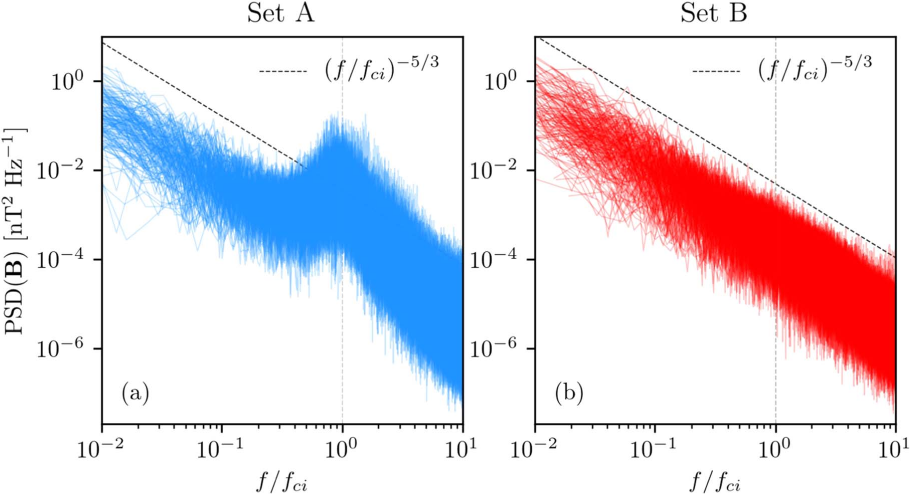}
\caption{Magnetic field power spectra density as a function of the observed normalized frequency for 30-minute events, seen by MAVEN MAG when is upstream and magnetically disconnected from the bow shock of Mars.  Set A (left) and Set B (right) consists of 184 events with PCW and 208 events without PCW, respectively. MAVEN: Mars Atmosphere and Volatile EvolutioN. MAG: Magnetometer. PCW: Proton Cyclotron Waves. Source:  Taken from \cite{Andres2020}.}
 \label{fig_andr_2020}
\end{figure}

Recent studies have focused on the effects that PCW have on Alfv\'enic turbulence in the pristine solar wind \citep{Ruhunusiri2017,Andres2020,Romanelli2022}. 
In particular, \cite{Andres2020} analyzed four months of MAVEN magnetic field and plasma data, 
to determine if the magnetic field power spectra and/or the energy cascade rates are affected by PCW. Figure \ref{fig_andr_2020} displays the magnetic field power spectral density (PSD) associated with 30-minute intervals in the pristine solar wind  \citep{Gruesbeck2018} when the IMF cone angle was approximately constant. All these events (sets A and B) have a normalized fluctuation plasma density  smaller than 20\%, thus corresponding to  nearly incompressible solar wind conditions. Set A contains cases with PCW, while these waves are absent in the events contained in Set B 
\citep{Andres2020}. As can be seen, Figure \ref{fig_andr_2020}, left panel, shows a clear peak in the PSD at the local proton cyclotron frequency for all events. These results also show a power law decay consistent with a Kolmogorov spectrum. Indeed, wave power can be modeled as $P \propto P_0 \, f^{-\gamma}$, where $f$ is the frequency and the spectral index, $\gamma=-5/3$, is constant in the MHD scales for all cases in both sets. The observed decay is consistent with  Alfv\'enic turbulence and does not appear to be affected by PCW at the MHD scales \citep{Andres2020}.

\cite{Andres2020} also estimated, for the first time, the absolute value of the incompressible energy cascade rate  of the solar wind upstream from Mars bow shock at MHD scales. 
For this, the authors made use of the exact relation for incompressible MHD turbulence \citep[see, e.g.,][]{P1998b,P1998a,Andres2016,Ferrand2021}.
Since MAVEN is a single spacecraft mission, \cite{Andres2020} assumed full isotropy and made use of the Taylor hypothesis to integrate Equation 4 in their work and compute the energy cascade rate per unit volume, $\varepsilon$ 
\citep[see, also,][]{St2011}.  Under these conditions, the isotropic nonlinear energy cascade rate can be expressed as a function of time lags $\tau$ as:

\begin{equation}\label{law_iso}
	\varepsilon = \rho_0\langle [(\Delta\uh\cdot\Delta\uh+\Delta\ua\cdot\Delta\ua)\Delta{u}_\ell - (\Delta\uh\cdot\Delta\ua+\Delta\ua\cdot\Delta\uh)\Delta{u}_{A\ell}]/(-4\tau U_0/3)\rangle.
\end{equation}

where $u_\ell={\bf u}\cdot{\bf \hat U}_0$,  $u_{A\ell}=\ua\cdot{\bf \hat U}_0$, and $U_0$ is the mean plasma flow speed. The angular bracket $\langle\cdot\rangle$ denotes an ensemble average \citep{Po2001}, which was taken as the time average assuming ergodicity. Note that   Equation \ref{law_iso} makes use of the increments definition,  $\Delta\alpha\equiv\alpha'-\alpha$, where 
fields are evaluated at position $\textbf{x}$ or $\textbf{x}'=\textbf{x}+\boldsymbol\ell$.  A prime is added to the corresponding field in the latter case. Equation \ref{law_iso} shows that under the considered assumptions, 
$\varepsilon$ is fully defined by velocity and magnetic field time increments, that \cite{Andres2020} computed based on MAVEN MAG and SWIA data.

\begin{figure}[ht]
\centering
\includegraphics[scale=0.65]{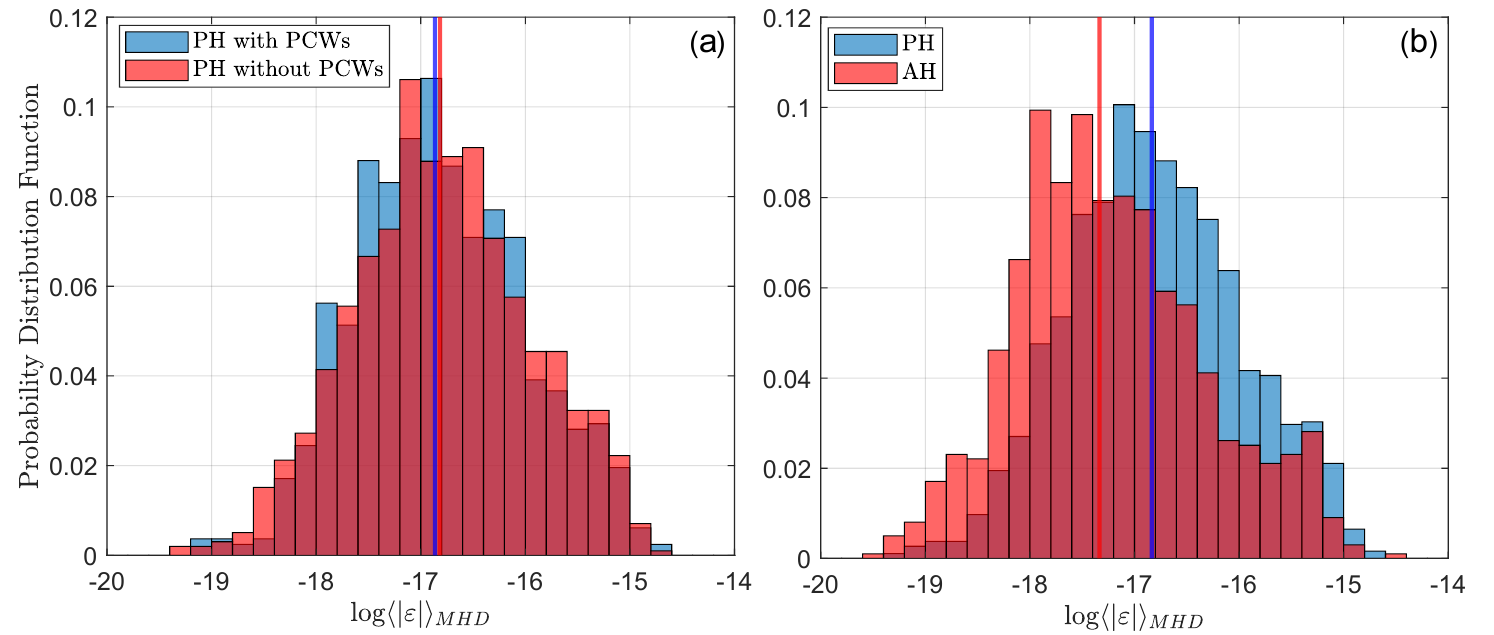}
\caption{Probability distribution function of log$\langle|\varepsilon|\rangle_{MHD}$ for different Martian orbital locations and PCW activity. (a) Martian perihelion with PCW (blue) and without PCW (red) (b) Martian perihelion conditions (with and without PCW, blue) and aphelion conditions (with and without PCW, red). (a-b) The vertical solid lines correspond to the median of the respective distributions. PCW: Proton Cyclotron Waves. Source: 
Taken from \cite{Romanelli2022}/ IOP Publishing/CC BY 4.0.}
 \label{fig_Romanelli_2022}
\end{figure}

\cite{Andres2020} reported an increase in the absolute value of the nonlinear solar wind incompressible energy cascade rate at MHD scales ($\langle|\varepsilon|\rangle_{MHD}$), when PCW are detected in the pristine solar wind. However, the known variability of PCW with the Martian heliocentric distance and the relatively small data set analyzed prevented the authors from determining the factor(s) responsible for such an increase.  
Making use of more than five years of MAVEN measurements and a similar methodology, \cite{Romanelli2022} concluded that $\langle|\varepsilon|\rangle_{MHD}$ displays a decreasing trend with the Martian heliocentric distance, in agreement with other studies \citep[e.g.,][]{H2017a,Ba2020, AN2021}. Moreover, the presented results suggest that PCW do not have a significant effect on 
$\langle|\varepsilon|\rangle_{MHD}$ \citep{Romanelli2022}. 
Figure \ref{fig_Romanelli_2022}a shows that the probability distribution function of log $\langle|\varepsilon|\rangle_{MHD}$ for conditions associated with Martian perihelion does not change significantly due to the presence of PCW. Both distributions are similar with log $\langle|\varepsilon|\rangle_{MHD}$  values ranging from $\sim-19$ to $\sim-15$, with $\langle|\varepsilon|\rangle_{MHD}$ median values equal to $1.4\times10^{-17}$ J m$^{-3}$ s$^{-1}$ and $1.5\times10^{-17}$ J m$^{-3}$ s$^{-1}$, for cases at perihelion with and without PCW, respectively. 
In contrast,  Figure \ref{fig_Romanelli_2022}b
displays a clear shift of the distribution with larger $\langle|\varepsilon|\rangle_{MHD}$  values around the Martian perihelion. The $\langle|\varepsilon|\rangle_{MHD}$ median value is equal to $1.5\times10^{-17}$ J m$^{-3}$ s$^{-1}$ and $4.6\times10^{-18}$ J m$^{-3}$ s$^{-1}$, for perihelion and aphelion conditions, respectively \citep{Romanelli2022}.

It is worth noticing that these results do not rule out the possibility of PCW effects on Alfv\'enic turbulence at kinetic scales.
 Complementary studies are needed to potentially identify these effects, using plasma observations with a higher sampling frequency and theoretical models of solar wind turbulence in this regime \citep{Ga2008,A2018}.  Moreover, future studies could also focus on  the effects that planetary foreshocks and associated wave activity have on solar wind turbulence. Planetary foreshocks are filled with electromagnetic plasma waves in different frequency ranges and can affect the nonlinear energy cascade rate in the MHD and kinetic scales \citep{Eastwood2005,Burgess2012}. 
The variability of solar wind properties and IMF strength and direction with heliocentric distance also offers the possibility to perform comparative analyses \citep[e.g.,][]{AndresNahuel2015,Meziane2017,R2018c,Romanelli2020waves,Romanelli2021,Jarvinen2022,Glass2023}. In this regard, numerical simulations also constitute a particularly useful approach to  understand better these  environments, improve data interpretation, and point towards fundamental physical processes taking place \citep[e.g.,][]{Jarvinen2022}.

\section{Alfv\'en waves in the Martian magnetosheath}

 Identifying wave modes in planetary magnetosheaths is particularly challenging for several reasons. Waves in the magnetosheath can be generated upstream and convected by the solar wind, {can be generated at the bow shock}, and/or be associated with local plasma instabilities {\citep[e.g.,][]{Dubinin2016,Harada2019}}. In the case of Mars, the latter also includes effects derived from the presence of planetary protons and oxygen ions  \citep[e.g.,][]{Chaffin2015,Clarke2017,Deighan2015,Feldman2011,Halekas2017seas,Barabash1991,Curry2015,Rahmati2015,Rahmati2017,Yamauchi2015,Dong2015}.  Furthermore, normal wave modes can be coupled in a nonuniform medium such as the magnetosheath \citep{Cramer2001}.
 
 It is also important to remark that conditions in the magnetosheath present asymmetries related to the local IMF and due to the deflection of the solar wind around Mars \citep{Dubinin2018,Romanelli2020deflection}. The former gives rise to temperature anisotropies, particularly downstream of the quasi-perpendicular bow shock \citep{Espley2004,Halekas2020,Wedlund2022,Jin2022,Wedlund2022egu}. The latter is responsible, for example, for superAlfv\'enic conditions in the Martian magnetosheath flanks, while the shocked solar wind is significantly decelerated near the subsolar MPB \citep{Halekas2020}. 

\begin{figure}[ht]
\centering
\includegraphics[scale=1.1]{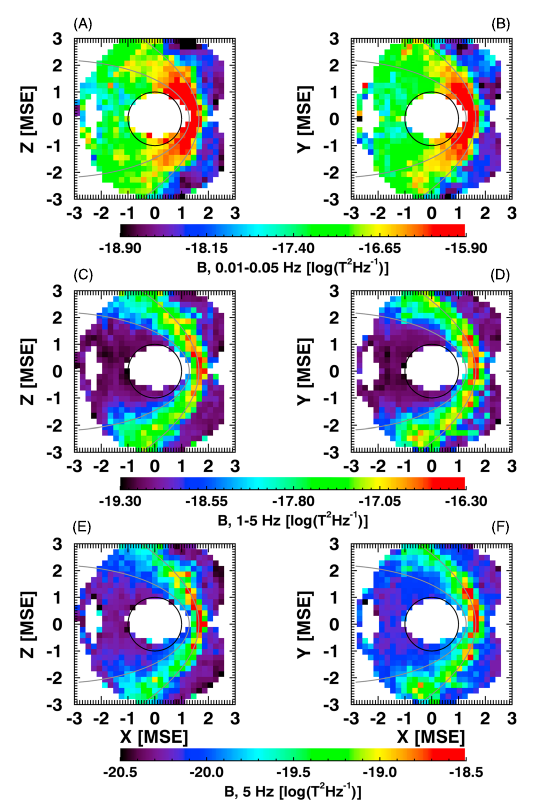}
\caption{Statistical magnetic field strength wave power for several frequency ranges, for altitudes greater than 600 km.
(a and b) 0.01–0.05 Hz, (c and d) 1–5 Hz, and (e and f) 5 Hz. The left and right columns correspond to projections onto the X-Z and X-Y MSE planes, respectively. MSE: Mars Solar Electric. Source:  \cite{Fowler2017}/ John Wiley \& Sons.}
 \label{fig_Fowler_2017}
\end{figure}

A detailed study on the electromagnetic wave power spatial distribution around Mars at different frequency ranges has been reported by \cite{Fowler2017}. The magnetosheath was found to be the region with the strongest magnetic field wave power in the ULF range \citep{Dubinin2016,Fowler2017}. This is associated with processes contributing to dissipating the bulk kinetic energy
 of the pristine solar wind  \citep[e.g.,][]{Papadopoulos1971,Auer1971,Burgess1989}. 
  Figure \ref{fig_Fowler_2017} left (right) column displays the magnetic field power distribution in the $X_{MSE}-Z_{MSE}$ ($X_{MSE}-Y_{MSE}$) plane. 
  
 Figures \ref{fig_Fowler_2017}a and \ref{fig_Fowler_2017}b show that most of the magnetic field wave power for frequencies between 0.01 Hz and 0.05 Hz lies in the magnetosheath, upstream of the terminator plane.  Magnetic field wave power is also significant in the magnetotail, although it is approximately one order of magnitude weaker throughout this region. The absence of a significant decrease in magnetic field ULF wave power between the magnetosheath and the ionosphere, shown in Figures \ref{fig_Fowler_2017}a and \ref{fig_Fowler_2017}b,
is consistent with the expected lack of complete ion thermalization in the magnetosheath.
Indeed, the relatively small size of the Martian magnetosheath {(its thickness along the Mars-Sun axis is $\sim$ 1200 km)} compared to ion spatial scales {(e.g., the convected solar wind proton gyroradius is $\sim$ 1000 km)} implies that kinetic effects are relevant and suggests that thermalization is likely incomplete {before plasma reaches the MPB} and inner regions of the Martian magnetosphere \citep{Moses1988,Kallio2011,Ruhunusiri2017,Jiang2023}. Thermalization processes involve space and time scales related to local ion gyrofrequencies, which in the Martian magnetosheath are near the ULF range \citep{Fowler2017}. 

Figures \ref{fig_Fowler_2017}c and \ref{fig_Fowler_2017}d  show that magnetic wave power in the 1-5 Hz range is maximum near the subsolar region and along the Martian bow shock. Wave power takes smaller values in the Martian magnetosheath and {is} three orders of magnitude smaller in the tail. Figures \ref{fig_Fowler_2017}e and \ref{fig_Fowler_2017}f show that wave power at 5 Hz presents a similar spatial distribution to that of between 1-5 Hz, but is two orders of magnitude smaller. Local maxima in the wave power at magnetospheric boundaries make evident that the solar wind kinetic energy is transferred to particle heating in these very thin regions. {Moreover, the associated particle velocity distribution function has potentially sufficient free energy for the local generation of electromagnetic waves.} Magnetic field wave power is also significant slightly upstream of the average bow shock location. This is due to several factors: the non-stationarity of this boundary, the presence of reflected ions at the shock, over and undershoots, and electromagnetic plasma waves. Among the latter, whistler waves, likely generated at the shock, are capable of propagating into and affecting the upstream region \citep{Trotignon2006,Halekas2017seas,Edberg2009a,Edberg2009b,Paschmann1981,Gosling1982,Eastwood2005,Mazelle2004,Livesey1982}.

So far,  the identification of wave modes in the Martian magnetosheath has been performed by computing transport ratios, interpreted in terms of MHD linear theory. 
Following the method firstly developed by \cite{Gary1993} and extended by \cite{Song1994}, 
\cite{Ruhunusiri2015} identified low-frequency wave modes as Alfv\'en and quasi-parallel slow waves (indistinguishable by the employed methodology), fast, quasi-perpendicular slow, and mirror mode waves. This was done based on four transport ratios:  the transverse ratio ($T_R$), the compressional ratio ($C_R$), the phase ratio ($P_R$) and the Doppler ratio ($D_R$). These are defined in terms of velocity, magnetic field, and density fluctuations as follows:

\begin{equation}
   T_R=\frac{\delta \mathbf{B} \cdot \delta \mathbf{B} - \delta B_{\parallel}^2}{\delta B_{\parallel}^2}
\end{equation}
\begin{equation}
   C_R=\frac{\delta n_{i}^2}{n_{i0}^2} / \frac{\delta \mathbf{B} \cdot \delta \mathbf{B}}{B_0^2}
\end{equation}
\begin{equation}
   P_R=\frac{\delta n_{i}}{n_{i0}}/ \frac{\delta B_{\parallel}}{B_0}
\end{equation}
\begin{equation}
   D_R=\frac{\delta \mathbf{u}_i \cdot \delta \mathbf{u}_i}{u_{i0}^2}/\frac{\delta \mathbf{B} \cdot \delta \mathbf{B}}{B_{0}^2}
\end{equation}
where $\delta B_{\parallel}$ is the  component of the magnetic field fluctuation ($\delta \mathbf{B}$) parallel to the background magnetic field $\mathbf{B_0}$. In addition, $\delta n_{i}$, $n_{i0}$, $\delta \mathbf{u}_i$, and $u_{i0}$ are the ion number density fluctuation, mean value, the fluctuation in the velocity field, and the mean ion velocity, respectively. The wave mode identification method developed by \cite{Song1994} assumes that a single wave mode is dominant at a single frequency with a single wavenumber. In addition, it is applicable in a high beta plasma, suggesting the results reported by \cite{Ruhunusiri2015} are most accurate in the Martian magnetosheath. {Table 1 in \cite{Ruhunusiri2015} reports the range of values for the transport ratios (Equations 5-8) used to identify low-frequency wave modes in the Martian magnetosphere.}

\begin{figure}[ht]
\centering
\includegraphics[scale=1.2]{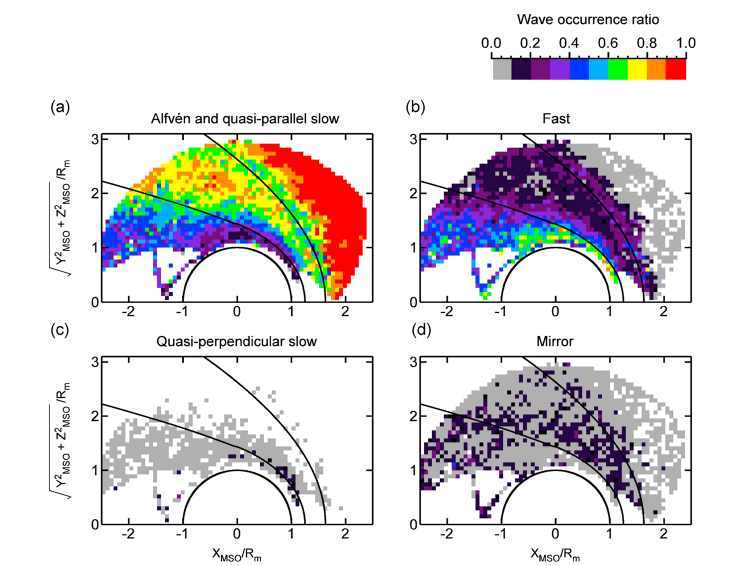}
\caption{Occurrence rate of several low-frequency wave modes. (a) Alfv\'en and quasi-parallel slow, (b) fast magnetosonic, (c) quasi-perpendicular slow, (d) mirror. Taken from \cite{Ruhunusiri2015} based on the identification technique developed in \cite{Song1994} for high beta plasmas.}
 \label{fig_Ruh_2015}
\end{figure}

Figure \ref{fig_Ruh_2015} displays maps of the wave mode occurrence ratio, based on MAVEN MAG, SWIA and Suprathermal and Thermal Ion Composition (STATIC) \citep{McFadden2015}  data between 7 October 2014 and 28 April 2015, in cylindrical MSO coordinates. \cite{Ruhunusiri2015} found a very high wave occurrence rate of Alfv\'en and quasi-parallel slow waves in the pristine solar wind and the magnetosheath (panel a). Several of the wave events upstream from the bow shock are likely Alfv\'en waves, in agreement with previous studies of upstream ULF waves and solar wind turbulence \citep[e.g.,][]{Andres2020,Halekas2020}. Moreover, the observed occurrence rate decrease from the upstream region into the magnetosheath may be indicative of upstream Alfv\'en wave convection. \cite{Ruhunusiri2015} also reported a significant amount of fast magnetosonic waves in the Martian magnetosheath that increases closer to the MPB (panel b). The source of these magnetosonic waves could be mode conversion at the bow shock or PCW and foreshock wave transmission from upstream to the magnetosheath. In addition, the authors also concluded there is a minor mirror mode wave population that is observed more frequently near the dayside magnetosheath region (panel d), in agreement with other magnetic field observations \citep{Espley2004,Wedlund2022,Wedlund2022egu}. Given that the wave identification method by \cite{Song1994} may not be applied downstream of the MPB (low beta plasma region) the authors did not provide final conclusions  about waves in this region \citep{Ruhunusiri2015}.

The previous analysis was not concerned with the nonlinear interaction between fluctuations of different temporal and spatial scales. \cite{Ruhunusiri2017} performed the first global characterization of turbulence in the Martian magnetosphere, based on the computation of spectral indices for perturbations observed in the magnetic field. As mentioned before, these indices are the slopes associated with the magnetic field power spectra distribution as a function of the observed frequency (in logarithmic scale). They provide information about the physical processes at play within a given frequency range (energy injection, transfer or cascade, and dissipation). 
In contrast with observations in the pristine solar wind, \cite{Ruhunusiri2017} concluded there is not an inertial range in the Martian magnetosheath following Kolmogorov's scaling (where $\gamma = -5/3$) expected when energy is transferred between scales.

\begin{figure}[ht]
\centering
\includegraphics[scale=1.3]{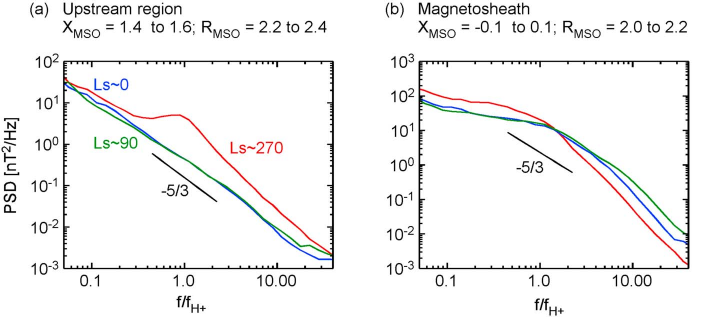}
\caption{Median magnetic field power spectra density as a function of the normalized frequency for three seasons in (a) the region upstream of the Martian bow shock and (b) the Martian magnetosheath. $f_{H^+}$ refers to the local proton cyclotron frequency. $L_s$ is the Martian solar longitude. Source:  \cite{Ruhunusiri2017}/ John Wiley \& Sons.}
 \label{fig_Ruh_2017}
\end{figure}

Figure \ref{fig_Ruh_2017} shows the median magnetic field power spectral density as a function of the normalized frequency for several Martian solar longitude ($L_s$) values.  $L_s$ is the angle between Mars and the Sun, measured from the northern hemisphere spring equinox where $L_s=0^\circ{}$. The Martian perihelion and aphelion correspond to $L_s=251^\circ{}$ and $L_s=71^\circ{}$, respectively. The northern hemisphere summer (winter) solstice occurs at  $L_s=90^\circ{}$ ($L_s=270^\circ{}$).
\cite{Ruhunusiri2017}  found that the spectral indices in the magnetosheath (panel b) take larger  values ($\gamma \sim -0.5$) for frequencies smaller than the local proton gyrofrequency and much more negative values ($\gamma \sim -2.7$) at higher frequencies.
This is in agreement with observations at other planetary magnetosheaths and suggests that the shocked solar wind, heated and slowed down at the bow shock, is composed of fluctuations that do not have sufficient time to interact nonlinearly and give rise to a fully developed energy cascade rate  \citep[e.g.,][]{Hadid2015,Zimbardo2010,Tao2015}. Although there seem to be some effects of PCW in the Martian magnetosheath spectra,  they do not appear as pronounced as in the region upstream of the bow shock (panel a). Note the clear peak at $f/f_{H^+} \sim 1$, for $L_s \sim 270^\circ{}$ when MAVEN is in the upstream region. In other words, the Martian magnetosheath appears to be filled with high amplitude Alfv\'en waves that have not reached a fully developed turbulent regime, at least near the terminator plane \citep{Ruhunusiri2015,Ruhunusiri2017,Fowler2017}.

\begin{figure}[ht!]
\centering
\includegraphics[scale=1.5]{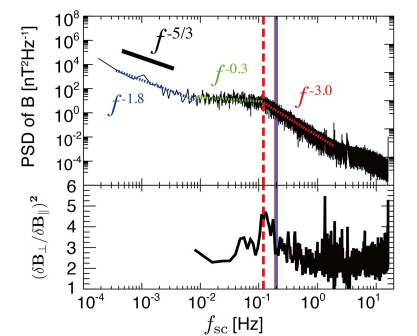}
\caption{(Upper) Magnetic field power spectral density for a Martian magnetosheath interval on 4 June 2016, between 21:10 UT and 22:43 UT. Vertical lines correspond to relevant ion-scale frequencies. The red dashed line shows the highest frequency breakpoint f$_{bk2}$= 0.12 Hz. The light red and light blue lines display f$_{\rho i}$ and f$_{d i}$, respectively, with f$_{\rho i} = V_i/2\pi \rho_i$, f$_{d i}  = V_i/2\pi d_i$, where $V_i$ is the local plasma bulk velocity, $\rho_i$ is the local thermal proton gyro-radius, and $d_i$ is the proton inertial length. (Lower) Square of the magnetic field perpendicular fluctuation normalized by the  magnetic field parallel fluctuation, as a function of the observed frequency in the spacecraft reference frame. Source: \cite{Jiang2023} / John Wiley \& Sons.}
 \label{fig_Jiang2023}
\end{figure}

Similar conclusions have also been reached by \cite{Jiang2023}. However, the authors also reported the presence of a Kolmogorov-like spectrum in the Martian magnetosheath for very low wave frequencies. The authors found magnetic field spectra in the Martian magnetosheath consisting of triple power laws, where the intermediate regime displays a plateau. An example of such events is shown in  Figure \ref{fig_Jiang2023}. Note that their breaking frequency points are typically between $\sim10^{-2}$ to $\sim10^{-1}$ and $\sim1$ to $\sim10$ times the local gyrofrequency, respectively, depending on different parameters, such as the plasma beta (see Figure 7c and 7d in that work). Their statistical analysis allows the identification of a significant correlation between 
the occurrence rate of plateau power spectra and planetary proton pick-up ion parameters. Specifically, the authors suggest the formation of plateau-like spectra may be due to energy injection associated with PCW \citep{Jiang2023}. It is possible, however, that a fully developed energy cascade rate is reached further downstream from the planet, along the magnetosheath flanks. Future studies may shed light on this matter. 

\section{Alfv\'en waves inside the MPB and the ionosphere}

While there have been studies of Alfv\'en waves in several regions of the Martian magnetosphere, there are far fewer studies focused on this phenomenon within the Martian MPB and  ionosphere. This is  likely due to a combination of factors such as a relatively low level of wave activity and the lack of adequate instrumentation and spacecraft sampling until more recent missions. 

\begin{figure}[ht]
\centering
\includegraphics[scale=1.5]{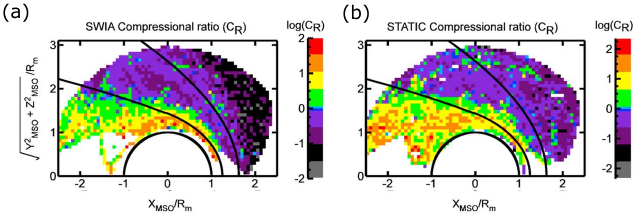}
\caption{Compressional ratio $C_R$ based on MAVEN MAG and SWIA (panel a) and STATIC (panel b) observations as a function of cylindrical MSO coordinates. MAVEN: Mars Atmosphere and Volatile EvolutioN. MAG: Magnetometer. SWIA: Solar Wind Ion Analyzer. STATIC: Suprathermal and Thermal Ion Composition. MSO: Mars Solar Orbital. Source: \cite{Ruhunusiri2015}/ John Wiley \& Sons.}
 \label{fig_Ruh_2015CR}
\end{figure}

Figure \ref{fig_Ruh_2015CR}  shows the compressional ratio $C_R$ is larger than 1 inside the Martian MPB. This result is observed both using MAVEN SWIA (panel a) and STATIC (panel b) data  and suggests Alfv\'en waves are not commonly present in the magnetic pile-up region and magnetotail \citep{Ruhunusiri2015}.  
In addition, as can be seen in Figure \ref{fig_Fowler_2017}, the magnetic field wave power decreases downstream of the MPB, particularly in the magnetotail. This could be due to relatively weak energy dissipation processes in the explored frequency ranges, as well as the much lower plasma beta values in this region \citep{Fowler2017}. 
The spectral indices in the magnetic pile-up region and the Martian wake are near $-2$ or smaller in both the MHD and kinetic ranges  \citep{Ruhunusiri2017}. The lack of a spectral break at the local proton cyclotron frequency may result from the plasma containing primarily heavy planetary ions instead of protons \citep{Ruhunusiri2017}. Previous reports have found similar results for the Venusian wake at frequencies between 0.01 and 0.5 Hz \citep{Voros2008a,Voros2008b}. 

The absence of Alfv\'en wave observations in the Martian ionosphere and/or along crustal magnetic fields may also be the result of limited instrumentation and spacecraft orbital geometry. 
 Early Mars orbiters, such as Phobos-2, carried magnetometers but did not sample the ionosphere. More recent missions such as MGS and MEX periodically sampled the ionosphere, but their ionospheric coverage was limited. While MGS MAG/Electron Reflectometer provided observations of the electromagnetic wave environment at Mars, the $\sim 400$ km circular orbit precluded any comprehensive study within the ionosphere. {In contrast,} MEX periapsis altitudes reach down to $\sim$275 km, providing in situ samples of the topside upper Martian ionosphere but it is unable to reach the main ionospheric peak, located at $\sim$125 km altitude at the sub-solar point.  Indeed, the main peak has been observed via remote sensing using the Mars Advanced Radar for Subsurface and Ionospheric Sounding radar on MEX  
 \citep[e.g.,][]{Gurnett2005,Dubinin2006,Orosei2015,Picardi2004}. Moreover, the lack of a magnetometer {onboard MEX} precludes any detailed study of the electromagnetic wave environment.  
	In the case of MAVEN, periapsis typically lies between 150 and 200 km, determined by targeting a specific neutral density corridor. MAVEN completed {nine} deep dip campaigns during the first few years of its mission, where periapsis altitude was lowered to $\sim$125 km for roughly week-long periods. This enabled MAVEN to sample down to the main ionospheric peak at mid to high solar zenith angles ($\sim$45$^\circ{}$ – 90$^\circ{}$). Also, MAVEN's precessing orbit provides comprehensive coverage throughout the magnetosphere and the Martian ionosphere. 
 
 The disparity in the study of Alfv\'en waves within the Martian ionosphere, compared to the rich and diverse collection of work related to their analysis within the terrestrial ionosphere, also likely lies in the underlying limitations of in-situ plasma measurements made at bodies other than Earth. In spite of particle and field measurements becoming more commonplace on interplanetary missions, their sampling frequency is typically much lower than their terrestrial counterparts. For example, the 
  ion (electron) distribution functions measured by the MEX Analyser of Space Plasma and Energetic Atoms 3 - Ion Mass Analyzer - (Electron Mass Spectrogram) are accumulated over 192 (4) seconds, and MAVEN STATIC and Solar Wind Electron Analyzer ion and electron distribution functions are  accumulated {over 4 and 2 seconds, respectively} \citep{Dubinin2006,McFadden2015,Mitchell2016}. Such measurements typically resolve the macro-scale parameters of the system, however, they are usually limited in resolving kinetic scale physics based on electromagnetic waves at frequencies of a few Hz up to 1000s Hz in the Martian plasma environment. In contrast, terrestrial space physics missions such as Cluster, Fast Auroral SnapshoT (FAST), and Magnetospheric Multiscale (MMS) measure the ion and electron distribution functions at tens of milliseconds \citep{Escoubet1997,Carlson1998,Burch2016}. The fields measurements (i.e. magnetic and electric fields) on interplanetary missions suffer from similar limitations: magnetic field data are typically limited to DC measurements at 32 Hz (compared to AC measurements of kHz at Earth), while only a single year of 1D electric field wave power spectra is available for Mars, made by the MAVEN  Langmuir Probe and Waves (LPW) instrument \citep{Fowler2017,Andersson2015}.   

{Recent studies have shown} electromagnetic waves {inside the MPB \citep[e.g.,][]{Harada2016}, in the ionosphere \citep[e.g.,][]{Gurnett2010,Esman2022}, and the Martian surface \citep[e.g.,][]{Johnson2020,Mittelholz2021,Mittelholz2023}; some of them} in connection with upstream processes leading to their generation \citep[e.g.,][]{Fowler2020,Wang2023,Shane2019,Shane2021,Shane2022}. These have been {primarily} whistler waves observed at the interaction region between the shocked solar wind and the upper ionosphere.  A comprehensive study on the occurrence rate and spatial distribution of these waves has not been undertaken yet. Next, we briefly review these works, {which} illustrate some of the potential effects Alfv\'en waves generated upstream may have in the Martian ionosphere.

\cite{Fowler2020} showed that whistler waves are produced in the upper ionosphere via a process known as 'magnetic pumping'. A summary of their event is shown in Figure 2 of their work. The free energy to drive the generation of the whistler waves is sourced from ULF compressive magnetosonic waves that are generated due to the Mars-solar wind interaction and that propagate across the draped magnetic field into the dayside ionosphere. Adiabatic compression of the plasma along wave fronts in the magnetic field acts to drive the electron distribution function unstable to the generation of whistler waves by creating an anisotropic distribution
\citep[e.g.,][]{Kennel1966}. The whistler waves were shown to break the reversibility of the underlying ULF magnetic pumping via efficient wave-particle interactions with the anisotropic electrons, leading to localized heating of this population. Such heating may play important roles in, for example, driving ionospheric photochemistry \citep{Fox2001}, the dissociative recombination of O$_2^+$ \citep{Lillis2017}, and the creation of ambipolar electric fields (-$\nabla p_e$) that can drive {ions} to escape to space \citep{Ergun2016}. Further study is needed to fully evaluate the impact on the energy and dynamics of the Martian ionosphere and to determine the occurrence rate of this process.

On the other hand, \cite{Wang2023} observed whistler waves within localized 'magnetic dips' at the interaction region between the solar wind and upper dayside ionosphere, in a manner postulated to be similar to whistler wave generation in the equatorial magnetosphere 
\citep[e.g.,][]{LeDocq1998}. The exact whistler generation mechanisms could not be determined, but temperature anisotropy and electron beam instabilities were proposed as likely candidates. The background plasma environment at Mars is such that these whistler waves propagate at much slower velocities than at Earth and so localized heating rather than propagation of the whistlers may be more important at Mars. Additional analysis is required to better understand the effects these waves have on the local plasma dynamics.

In addition, a comprehensive series of papers by \cite{Shane2019,Shane2021,Shane2022}
 used numerical approaches to demonstrate that whistler wave-particle interactions with suprathermal electrons can explain observed pitch angle distributions (PADs) on closed crustal field lines which intersect the Martian ionosphere. Many previous studies have shown that $>$100 eV PADs on dayside closed field lines are isotropic or trapped, however, these PADs are not expected based on collisional scattering and conservation of adiabatic invariants alone (which are important as the electrons mirror within the closed crustal field loops) {\citep[e.g.,][]{Liemohn2003,Brain2007}.} 
	Numerical calculations of the bounce-averaged electron diffusion coefficients show that wave-particle interactions are more efficient than Coulomb collisions above the exobase region and can play an important role in shaping the PADs in a manner consistent with observations \citep{Shane2021}. Numerical simulation of the bounce-averaged electron diffusion equation, including whistler wave resonance, demonstrated that such whistler wave-particle interactions alter the PADs and are consistent with observations \citep{Shane2022}.  The authors noted that future work should determine whether the frequency of whistler waves is high enough to explain the average behavior in observed PADs, and determine how such whistler waves would be generated.

Several studies have also shown that solar wind pressure pulses  can be convected into the bow shock and 'ring' the planetary magnetosphere \citep[e.g.,][]{Collinson2018}. Pressure pulses propagate through the magnetosphere in the form of fast magnetosonic waves (traveling across the draped magnetic field), where they can reach both the day and nightside ionosphere and drive heating of the light and heavy ion species through a variety of wave-particle interaction processes \citep{Fowler2018b,Fowler2021}.

A final note should be made with respect to the localized crustal magnetic field regions at Mars. These regions act as 'mini-magnetospheres' that rotate with the planet, and can sporadically magnetically reconnect with the solar wind IMF 
\citep[e.g.,][]{Weber2020,Bowers2023} to create 'mini cusp like' regions. Given the similarity of these cusp regions with the auroral zones at Earth, one may expect electromagnetic waves to play similar crucial roles in the physical processes present in these regions (primarily precipitation and auroral physics \citep{Andre1997}). In particular, \cite{Ergun2006} performed numerical calculations to show that electromagnetic waves associated with the Mars-solar wind interaction and field-aligned currents in regions of crustal magnetic fields could heat ionospheric O$^+$ and O$_2^+$ to escape energy and drive escape rates of $\sim10^{25}$ s$^{-1}$, via cyclotron damping. Their predictions closely match total ion escape rates that have been calculated based on recent observations in the magnetotail region of Mars \citep[e.g.,][]{Dong2015,Ramstad2015,Brain2015}.
 However, the full range of ion energization mechanisms has not been comprehensively identified or quantified. There are still many open questions about the initial acceleration of ions in the Mars ionosphere, and the role that electromagnetic waves play in it \citep{Hanley2022}. {In addition, a full 3D electric field in-situ instrumentation at low and medium frequencies, together with magnetic field and particle measurements, are needed 
 in order to identify the exact wave modes taking place in these regions.}

 \section{Conclusions}

Over the last few years, there has been significant progress in characterizing the Alfv\'enic input and content of the Martian magnetosphere. The pristine solar wind, upstream of Mars's  bow shock, has been characterized in terms of Alfv\'enic turbulence, where the absolute value of the incompressible nonlinear energy cascade rate is on the order of  $10^{-17}$ J m$^{-3}$ s$^{-1}$ at MHD scales. Studies have also shown that the cascade rate varies with Mars' heliocentric distance, in agreement with investigations focused on the solar wind closer to the Sun and upstream of other planetary magnetospheres \citep{Andres2020,Romanelli2022,H2017a,Ba2020, AN2021}. Moreover, values of normalized cross-helicity and residual energy show that the solar wind presents more Alfv\'en waves propagating outwards from the Sun (than inwards) and that the magnetic field fluctuations have more energy than the kinetic field counterpart \citep{Andres2020,Halekas2017}.
PCW associated with the extended hydrogen corona
do not appear to modify the solar wind energy cascade rate at MHD scales, although their energy could have potential effects on the kinetic regime \citep{Romanelli2022}. 
The magnetosheath is characterized by high amplitude wave activity, and a high occurrence rate of Alfv\'en waves, although other wave modes are also present \citep{Ruhunusiri2015,Fowler2017}. 
The large amplitude waves are likely the result of (incomplete) thermalization processes of the solar wind. The relatively small size of the magnetosheath, together with potential effects from the bow shock, PCW, and other wave modes are also responsible for a magnetic field turbulent spectrum that is not fully developed \citep{Fowler2017,Jiang2023}. Among the secondary wave populations in the magnetosheath, there exists a significant component  of fast magnetosonic waves and a minor contribution of mirror mode waves, particularly in the dayside \citep{Ruhunusiri2015,Espley2004,Wedlund2022egu}. Downstream of the magnetic pile-up boundary, waves generally display less magnetic field power \citep{Fowler2017}. The PSD of the magnetic field does not show a Kolmogorov-like spectrum but rather is characterized by a spectral index more negative both below and above the local proton gyrofrequency \citep{Ruhunusiri2017}.

Early studies of the solar wind interaction with Mars predicted that significant electromagnetic wave activity would be present at the interface, and within, the Martian ionosphere, as a direct result of the relatively small scale size of the magnetospheric system \citep{Ergun2006,Kallio2011}. Subsequent studies have confirmed this and hint that a plethora of electromagnetic wave modes exist and interact with the ionospheric plasma to drive energization and dynamics \citep[e.g.,][]{Fowler2018,Fowler2021,Collinson2018}. Several of these studies have identified the presence of whistler mode waves and subsequent wave-particle interactions in this Mars-solar wind interaction region. However, Alfv\'en waves have not been conclusively detected in the ionosphere or along remanent crustal magnetic fields yet. This could be due to the instrumental cadence or orbital geometry associated with previous missions to Mars as well as the lack of consistent conditions leading to their excitation.

Despite this progress, there is still much to be done to definitively identify wave modes and the processes giving rise to them at Mars. The lack of sufficient studies is partly due to the relatively small amount of simultaneous ion plasma and magnetic field data before the current MAVEN mission and the dynamics behavior of the Martian environment, among other factors. Mars possesses a hybrid magnetosphere with elements of induced and intrinsic magnetospheres \citep{Mazelle2004,Brain2003,DiBraccio2022,Dubinin2023}. The extended hydrogen {corona} is also responsible for additional processes and their variability {over} long (seasonal) timescales \citep{Halekas2017seas,Romanelli2016,Romeo2021}. Also, the use of single spacecraft observations does not allow computing the Doppler shift {for low-frequency waves}, due to the relative motion between the plasma rest frame and the spacecraft reference frame. Given that the solar wind velocity is close to or larger than the phase speed of many waves inside and upstream of the Martian magnetosphere, the waves are observed at different speeds and, in some cases, with different polarization than that seen in the plasma reference frame \citep{Brinca1991}.
As a result, so far many studies on waves have investigated the conditions that favor their excitation and basic properties, such as the observed frequency and polarization, as opposed to looking at the intrinsic wave frequency and wavelength and determining the corresponding wave mode.

Future studies could be focused on the determination of the direction of the energy cascade rate upstream of Mars and the effects that PCW and foreshock waves have on the MHD and kinetic regimes. Comparison between studies at different planetary foreshocks is also an interesting venue for future work. Moreover, an analysis of the processes occurring in the Martian bow shock and magnetic pile-up boundary would be beneficial to better understand the nature of the wave activity present inside the Martian magnetosphere, in particular the magnetosheath. Additional unknowns remain including wave mode identification downstream of the MPB and Alfv\'en wave detection near the ionosphere or along the crustal magnetic fields. If the latter are present, additional analyses should be focused on determining their occurrence rates  (in time and space), their production mechanisms, their impacts on the local plasma, and their impact on more global parameters.

 A combination of current in-situ measurements and global simulations can address some of these unknowns. Hybrid numerical simulations are particularly useful to improve data analysis and point out physical processes potentially taking place in the Martian magnetosphere \citep{Jarvinen2022,Modolo2016,romanelli2018cme}.
 They are also important as 
  wave amplitudes are often large, suggesting linear wave theory is limited (if not invalid) in interpreting spacecraft observations. 
  However, answers to some of the previous questions will likely remain elusive until {measurements
equivalent in capability to those made by terrestrial space physics missions
} can be made at Mars. 
{In particular, simultaneous multi-spacecraft and multi-probe electric and magnetic field measurements with high-time resolution particle instruments are needed to determine the exact wave modes, polarization, and Poynting flux. These observations would allow, in turn, the determination of the growth and dissipation of waves in various regions in the Martian magnetosphere.  In this regard,} multi-spacecraft missions such as the planned Escape and Plasma Acceleration and Dynamics Explorers (ESCAPADE)  will certainly contribute to improving our current understanding of waves in this magnetosphere \citep{Lillis2022}.

\section*{Acknowledgements}
\label{sec:acknowledgements}

The MAVEN project is supported by NASA through the Mars Exploration Program. The material is based upon work supported by NASA under award number 80GSFC21M0002. Support for this research was also provided by GSFC/EIMM within
NASA’s Planetary Science Division Research Program.


\end{document}